\documentclass[aps,pra,preprintnumbers,showpacs,a4paper,twoside,twocolumn,floatfix,10pt]{revtex4-1}

\usepackage{graphicx}
\usepackage{amsmath}
\usepackage{amssymb}
\usepackage{amsthm}
\usepackage{amsfonts}
\usepackage{amsbsy}
\usepackage{bm}
\usepackage{bbm}
\usepackage{setspace}
\usepackage{topcapt}
\usepackage{color}
\usepackage[thinspace,amssymb]{SIunits}
\usepackage[plainpages=false]{hyperref}
\hypersetup{
  pdfauthor={Heike Schwager and J. Ignacio Cirac and Geza Giedke},
  pdftitle={Dissipative spin chains: Implementation with cold atoms and steady-state properties}
  pdfsubject={},
  urlcolor=blue,
}
\newcommand{\bra}[1]{\langle #1 \bigr\rvert}
\newcommand{\ket}[1]{\bigl\lvert#1\rangle }
 
\newcommand{\kla}[1]{\left( #1 \right)}
\newcommand{\klasmall}[1]{\big( #1 \big)}

\renewcommand{\Re}{\mathrm{Re}}
\renewcommand{\Im}{\mathrm{Im}}
\newcommand{\mr}[1]{\mathrm{#1}}
\newcommand{\id}{\ensuremath{\mathbbm{1}}} 
\newcommand{\bcH}{\pmb{\mathcal{H}}}
\newcommand{\bcL}{\pmb{\mathcal{L}}}
\newcommand{\bcS}{\pmb{\mathcal{S}}}
\newcommand{\bcU}{\pmb{\mathcal{U}}}
\newcommand{\bcV}{\pmb{\mathcal{V}}}

\begin{document}

\title{Dissipative Dynamics and Phase Transitions in Fermionic Systems}

\author{Birger Horstmann$^{1,2}$}
\author{J. Ignacio Cirac$^1$}
\author{G\'eza Giedke$^{1,3}$}
\affiliation{(1) Max-Planck-Institut f\"ur Quantenoptik, Hans-Kopfermann-Stra\ss e 1, 85748 Garching, Germany}
\affiliation{(2) Deutsches Zentrum f\"ur Luft- und Raumfahrt, Institut f\"ur
  Technische Thermodynamik, Pfaffenwaldring 38-40, 70569 Stuttgart, Germany}
\affiliation{(3) Zentrum Mathematik, Technische Universit\"at M\"unchen,
  L.-Boltzmannstr. 3, 85748 Garching, Germany}
\date{Feb 15, 2013}

\begin{abstract}
  We study abrupt changes in the dynamics and/or steady state of fermionic
  dissipative systems produced by small changes of the system parameters.
  Specifically, we consider fermionic systems whose dynamics is described
  by master equations that are quadratic (and, under certain conditions,
  quartic) in creation and annihilation operators. We analyze phase
  transitions in the steady state as well as ``dynamical transitions''. The
  latter are characterized by abrupt changes in the rate at which the system
  asymptotically approaches the steady state. We illustrate our general
  findings with relevant examples of fermionic (and, equivalently, spin)
  systems, and show that they can be realized in ion chains.
\end{abstract}

\maketitle

\section{Introduction}

Motivated by the impressive experimental control over many-body
quantum states and dynamics \cite{Southwell08}, open many-body quantum
systems have received increasing experimental and theoretical
attention in recent years.  On the one hand, the decoherence
introduced by coupling to an environment is a major challenge to
quantum information processing \cite{Shor95}, on the other hand, it
can play a constructive role for quantum computing
\cite{Verstraete2009, Kraus2008}, state preparation
\cite{Diehl2008,Diehl10b,Roncaglia10,Yi2012}, entanglement generation
\cite{Muschik10,Krauter10}, quantum memories \cite{Pastawski11} or
quantum simulation
\cite{Syassen08,Durr09,GarciaRipoll2009,Daley09,KiHa10}.

These exciting possibilities drive the interest in understanding the steady-state phase diagram of open systems in detail \cite{Diehl2010}.  Of particular
interest are points of transitions between different phases of the system. For
closed systems at zero temperature, the phase diagram and quantum phase
transition can be understood by studying the low-lying energy eigenstates of
the system's Hamiltonian \cite{Sachdev}. In particular, the non-analyticity of
certain expectation values as a function of an external parameter, that
characterizes the quantum phase transition, can only occur if the gap of the
Hamiltonian closes, i.e., the energy difference between ground state and first
excited state vanishes. Quantum phase transitions are thus determined by the
low energy spectrum of the Hamiltonian governing the dynamics of wave
functions
\begin{equation}
\partial_t\ket{\Phi}=-\frac{i}{\hbar}\mathbf{H}\ket{\Phi}.
\end{equation}

In this paper we study abrupt changes in the physical properties of a
many-body quantum system whose dynamics is described by a master equation
\begin{equation}\label{eq:Meq}
 \partial_t\rho=\bcS\rho.
\end{equation}
This equation describes the dynamics of an open system coupled to a Markovian
reservoir \footnote{%
For most of this work, we take the Lindblad master
  equation as given and are not concerned with its microscopic derivation from
a particular coupling to some environment.}, where $\rho$ is the system's density operator. The superoperator
$\bcS$ contains two parts: one is related to the system
Hamiltonian (eventually renormalized due to the interaction with the
environment) and the other to the dissipation induced by the
environment. Under the appropriate conditions, the system evolves to a steady
state $\rho_\mr{ss}$, which corresponds to a (right) eigenstate of $\bcS$ with
eigenvalue 0. Note that this eigenvalue may be degenerate, or there may be
other eigenvalues with zero real part. In case this does not happen, the steady state is unique. Then, the
other eigenvalues $\lambda$ of $\bcS$ have a negative real part, and the
smallest absolute value of them, $\Delta$, determines the \emph{asymptotic
  decay rate} (ADR), that is, the rate at which the steady state is reached. A
phase transition in the steady state, where its properties abruptly change
when one slightly changes a parameter in the master equation will be
accompanied by the vanishing of $\Delta$. 
This situation has been studied by
many authors recently (see, for example,
\cite{Diehl2008,Verstraete2009,Diehl2010,EiPr11,Kessler2012a}) and might be 
referred to as a ``dissipative quantum phase transition''. There is a
natural analogy between dissipative and (closed-system) quantum phase
transitions: A unique ground state of the Hamiltonian is analogous to a unique
steady state. The appearance of a phase transition is signaled by the
vanishing of the gap or $\Delta$, respectively.

Apart from its role in reflecting the appearance of a phase transition, the
quantity $\Delta$ can play an additional role. It also represents a physical
property of the system, namely the rate at which the steady state is
approached asymptotically or the system's response to perturbations in the steady
state. This quantity may change abruptly itself. In that case, we can talk
about a \emph{dynamical transition}, since a small change in the system
parameters may lead to an abrupt change of the dynamics of the
system. Actually, such a transition may in principle occur even if $\Delta$
remains finite, and thus it is a different property than the transitions
generally studied in this context.

In this paper we investigate both kinds of transitions for simple
fermionic systems.  We concentrate on systems that are
  described by master equations in which the
Hamiltonian part is at most quadratic in fermionic creation and
annihilation operators.  Additionally, we consider two kind of
dissipative parts in terms of their dependence on such operators: (i)
general quadratic and (ii) quartic, but with some conditions (in
particular, that they correspond to Hermitian Lindblad operators). In
the first case, the dynamics can be exactly solved
\cite{Prosen08,Prosen10,PrZu10,Hartmann10} which has been exploited in
several recent works to study the interplay of dissipation and
critical Hamiltonians in 1d fermionic systems
\cite{PrZu10,EiPr11,HMF12}. In the second case, even though the full
dynamics cannot be obtained, we will show that it is nevertheless
possible to exactly determine the dynamics of certain expectation
values, from which dynamical and steady-state properties can be
obtained. In this last case we will present analytical examples where
dynamical transitions occur \cite{Horstmann2011}. This situation has
also been studied in \cite{Znidaric10,Znidaric11,Eisler11} with
particular regard to transport through a dephasing spin chain, where
exact solutions of the associated master equation could be obtained.

The formalism we develop is relatively general and we illustrate it with
explicit examples. In particular, we consider Hamiltonians which are
intimately connected to physical situations that can be obtained in the lab,
namely anisotropic $XY$ spin chains in transverse magnetic fields, and that
are mapped to a fermionic Hamiltonian by a Jordan-Wigner transformation. This
family of Hamiltonians displays the prototype of a continuous phase transition
\cite{Sachdev}. The dissipative terms we consider can also be understood as
particular physical processes occuring in the spin chain through its
interaction with an environment \cite{Zhao2012}.  Note that our framework also applies to the
systems studied in \cite{Prosen08,Prosen10,Znidaric10,Znidaric11}, and for the
quadratic dissipative terms is related to \cite{PrZu10,EiPr11}, where generic
dissipative phase transitions are analyzed.

This paper is structured as follows. In Sec.~\ref{system} we introduce the
Lindblad master equation which allows to describe decoherence due to the weak
interaction with a Markovian bath and present the covariance matrix formalism,
which allows the exact treatment of quadratic fermionic systems. In
Sec.~\ref{Formalism} we extend this formalism to decoherent systems with
linear and Hermitian quadratic Lindblad operators. Then we come to the
calculation of the steady states and the ADRs for relevant interesting
examples in this framework in Secs.~\ref{sec:applic:linear},
\ref{sec:applic:quadratic}, and \ref{section XY chain}. Here we explicitly
demonstrate the presence of dissipative phase transitions. In
Sec.~\ref{Experiment} we propose a possible implementation with cold ions
before concluding in Sec.~\ref{Conclusion}.

\section{Notation and Methods}
\label{system}
In this section we introduce our tools and notation, namely the Lindblad
master equation and the fermionic covariance matrix (CM) formalism which is ideally
suited for describing quasi-free fermionic systems (see
Sec.~\ref{sec.covariance}).

\subsection{Lindblad Master Equation}
\label{sec.lindblad}
We consider systems whose interaction with an environment
  leads to a time-evolution governed by a 
Lindblad master equation \cite{Lin76a}
\begin{align}
\label{Lind1}
 \partial_t \mathbf{\rho}&=\bcS\rho\nonumber
\\&=-\frac{i}{\hbar}\left[\mathbf{H},\mathbf{\rho}\right]+\sum_\alpha \left(\mathbf{L}^\alpha\mathbf{\rho}\mathbf{L}^{\alpha\dagger}-\frac{1}{2}\left\{\mathbf{L}^{\alpha\dagger}\mathbf{L}^{\alpha},\mathbf{\rho}\right\}\right),
\end{align}
where $\rho$ is the density matrix of the system, $\mathbf{H}$ is its
Hamiltonian, and the Lindblad operators $\mathbf{L}^\alpha$ determine the
interaction between the system and the bath. This dynamical equation for an
open system can be derived from two different points of view
\cite{Breuer}: First, it can be derived from the full dynamics of system and
bath. Here three major approximation have to be used: The states of system and
environment are initially uncorrelated, the coupling between system and bath
is weak (Born approximation), and the environment equilibrates fast (Markov
approximation). 
Second, any time-evolution given by a quantum dynamical
semigroup (i.e., a family of completely positive, trace preserving maps
$\epsilon_t$, which is strongly continuous and satisfies
$\epsilon_t\epsilon_s=\epsilon_{t+s}$) is generated by an equation of the form
Eq.~(\ref{Lind1}).

We characterize the decoherence dynamics with the steady state and the
ADR. A steady-state density matrix $\rho_0$ of the
master equation \eqref{Lind1} fulfills
\begin{equation}
 \label{definition steadystate}
\partial_t\rho_0=\bcS\rho_0=0
\end{equation}
and is the (generically unique) eigenvector with eigenvalue 0 of the
Liouvillian superoperator $\bcS$. The approach to the steady
state is then governed by the non-zero eigenvalues (and eigenvectors) of
$\bcS$, all of which have non-positive real part for Liouvillians
of Lindblad form. Of particular interest is the eigenvalue with the largest
real part (i.e., smallest modulus of the real part), since it governs the long-term dynamics. We refer to the absolute
value of this largest real part as the ADR and
denote it by $\Delta$:
\begin{equation}
  \label{eq:def-ADR}
\Delta(\bcS) = \mr{max}\{|\Re\lambda|\not=0 : \exists\rho_\lambda:
\bcS(\rho_\lambda)=\lambda\rho_\lambda\}.   
\end{equation}

\subsection{Quasifree Fermions and Spins}
\label{sec:system}

We consider systems with $N$ fermionic modes described by creation and
annihilation operators $a_j^\dagger$ and $a_j$. These operators obey the
canonical anti-commutation relations
\begin{equation}\label{eq:car1}
 \{a_j,a_k\}=0,\hspace{0.2 cm}
 \{a_j^\dagger,a_k\}=\delta_{jk}. 
\end{equation}
Equivalently, we can use Hermitian fermionic Majorana operators
\begin{equation}
 c_{j,0}=a_j^\dagger+a_j,\hspace{0.2 cm} c_{j,1}=(-i)\klasmall{a_j^\dagger-a_j},
\end{equation}
which as generators of the Clifford algebra satisfy the anti-commutation
relations
\begin{equation}\label{CAR}
\left\{c_{j,u},c_{k,v}\right\}=2\delta_{jk}\delta_{uv}.
\end{equation}

We consider fermionic Hamiltonians that are quadratic in the Majorana
operators. They describe quasifree fermions and are known to be exactly
solvable. We parameterize them with the real antisymmetric matrix $H$
\begin{equation}
\label{eq:HamMatrix}
 \mathbf{H}=\frac{i}{4}\hbar\sum_{jkuv}H_{jk,uv}c_{j,u}c_{k,v}.
\end{equation}
The $2\times2$ matrix $H_{jk}\equiv(H_{jk,uv})_{uv}$ describes the coupling
between the modes $j$ and $k$.

All eigenstates and thermal states of such a quadratic fermionic Hamiltonian
are Gaussian, i.e., they have a density operator which is the exponential of a
quadratic form in the Majorana operators. Gaussian states remain Gaussian
under the evolution with quadratic Hamiltonians.

In the following, we will mostly concerned with \emph{translationally invariant}
systems and nearest-neighbor interactions. In terms of the matrix $H$ the
former means that $H_{jk}$ depends only on the difference $j-k$ and we write
for short 
\begin{equation}
  \label{eq:HamTI}
  H_{jk}\equiv H_{j-k},
\end{equation}
 while the latter implies that $H_s=0$ for
$s>1$. We work with periodic boundary conditions, so $j-k$ is understood
modulo $N$. 

An important reason to study one-dimensional fermionic systems with quadratic
Hamiltonian is their intimate relation to certain types of spin chains: The
Jordan-Wigner transformation \cite{Nielsen2005} maps 
fermionic operators onto Pauli spin operators via
\begin{equation}\label{eq:JW}
 c_{j,0}\leftrightarrow\prod_{k=1}^{j-1}\sigma_z^k\sigma_x^j,\hspace{0.5cm}
 c_{j,1}\leftrightarrow\prod_{k=1}^{j-1}\sigma_z^k\sigma_y^j.
\end{equation}
Under this transformation some spin chains are mapped to spinless quasifree
fermionic systems which can be solved exactly. A prominent example is the
anisotropic XY chain in a transverse magnetic field \cite{Sachdev} with the
Hamiltonian
\begin{multline}
\label{XY chain}
\mathbf{H}=-J\sum_{j=1}^N\left[\klasmall{1+\gamma}\sigma_x^j\sigma_x^{j+1} +
  \klasmall{1-\gamma}\sigma^j_y\sigma_y^{j+1}\right]\\ 
+ B\sum_{j=1}^N\sigma^j_z,
\end{multline}
where $B$ is the magnetic field, $J$ the ferromagnetic coupling, and $\gamma$
the anisotropy parameter. Closed systems governed by this Hamiltonian show a
quantum phase transition at $B=2J$ in the thermodynamic limit and the
behavior in the presence of dissipation is studied in Sec.~\ref{sec.example}.

We are interested in dissipative (open) fermionic systems, with dynamics
described by a Lindblad master equation, characterized by a set of Lindblad
operators $L^\alpha$. We consider two classes of Lindblad operators: firstly,
those given by arbitrary linear combinations of the Majorana operators
(\emph{linear} Lindblad operators) 
\begin{equation}
\label{definition linear}
\mathbf{L}^\alpha=\sum_{ju}L^\alpha_{j,u}c_{j,u},\,\,\, L^\alpha_{j,u}\in\mathbb{C},
\end{equation}
and secondly, those represented by quadratic expressions in the Majorana
operators which are in addition Hermitian (\emph{Hermitian quadratic} Lindblad
operators)   
\begin{equation}
\label{definition quadratic}
\mathbf{L}^\alpha=\frac{i}{4}\sum_{jkuv}L^\alpha_{jk,uv}c_{j,u}c_{k,v}
\end{equation}
with the real and antisymmetric matrix $L^\alpha$.

\subsection{Covariance Matrix Formalism}
\label{sec.covariance}

Now we present a framework in which the dissipative dynamics of the Lindblad
master equation \eqref{Lind1} can be solved exactly. 

For every state of a fermionic system, its real and antisymmetric CM is defined by 
\begin{equation}
\label{covariance_matrix}
\Gamma_{jk,uv}={\rm tr}\kla{\rho\frac{i}{2}\left[c_{j,u},c_{k,v}\right]}.
\end{equation}
The magnitudes of the imaginary eigenvalues of $\Gamma$ are smaller than or
equal to unity ($\Gamma^2\leq-\mathbbm{1}$). 

For Gaussian states the correlation functions of all orders are related to the
CM through Wick's theorem \cite{Wick50}. In particular, pure
Gaussian states $\mathbf\rho=\ket\Psi\bra\Psi$ satisfy 
$\Gamma^2=-\mathbbm{1}$. In our notation $\Gamma_{jk}$ denotes a $2\times2$
matrix that describes the covariances between sites $j$ and $k$.

\section{Lindblad Master Equation in the Covariance Matrix Formalism}
\label{Formalism}
The CM formalism is especially useful if the operative dynamics
leads to closed equations for the CM, which is the case for the two kinds of
Lindblad operators Eqs.~(\ref{definition linear},\ref{definition quadratic})
that we study in the following. 

\subsection{Linear Lindblad operators}
\label{linear}
We consider a system with quadratic Hamiltonian given by the antisymmetric
matrix $H$ [cf.\ Eq.~\eqref{eq:HamMatrix}] and linear Lindblad operators as
defined in Eq.~\eqref{definition linear}. Using the anti-commutation relations
\eqref{CAR} we determine the dynamical equation for the CM
$\Gamma$ from Eq.~\eqref{Lind1} and obtain:
\begin{multline}
\label{Lindblad_linear}
\partial_t \Gamma=\left[H,\Gamma\right]-\sum_\alpha\left\{\ket{L^\alpha}\bra{L^\alpha}+\ket{L^{\alpha *}}\bra{L^{\alpha *}},\Gamma\right\}\\-2i\kla{\ket{L^\alpha}\bra{L^\alpha}-\ket{L^{\alpha *}}\bra{L^{\alpha *}}},
\end{multline}
where $\ket{L^\alpha}$ denotes the vector formed by the coefficients
$L^\alpha_{j,u}$ in Eq.~(\ref{definition linear}) and $\ket{L^{\alpha *}}$ its
complex conjugate. In terms of $\ket{\Gamma}$, the vector of components of
$\Gamma$, this equation becomes
\begin{equation}
\label{Lindblad_linear2}
\partial_t \ket{\Gamma}=\mathcal{S}\ket\Gamma-\ket{\mathcal{V}}=\kla{\mathcal{H}-\mathcal{M}}\ket\Gamma-\ket{\mathcal{V}},
\end{equation}
with the superoperators
\begin{align}
\label{superoperator_linear}
\mathcal{H}&=\kla{H\otimes\id-\id\otimes H^\text{T}},\\
\mathcal{M}&=\sum_\alpha\kla{\ket{L^\alpha}\bra{L^\alpha}\otimes\id+\id\otimes(\ket{L^\alpha}\bra{L^\alpha})^T+\text{c.c.}},\\
\ket{\mathcal{V}}&=2i\sum_\alpha\kla{\ket{L^\alpha}\otimes\ket{L^\alpha}-\text{c.c.}}.
\end{align}
Note that $\mathcal{H}$ is anti-Hermitian and $\mathcal{M}$ is Hermitian and
positive semi-definite. The steady-state CM [see
Eq.~\eqref{definition steadystate}] satisfies
\begin{equation}
\kla{\mathcal{H}-\mathcal{M}}\ket{\Gamma_0}=\ket{\mathcal{V}}.
\end{equation}
Deviations $\ket{\delta\Gamma}=\ket{\Gamma}-\ket{\Gamma_0}$ then obey 
\begin{equation}
  \label{eq:1}
 \partial_t \ket{\delta\Gamma}= \kla{\mathcal{H}-\mathcal{M}}\ket{\delta\Gamma}
\end{equation}
and the approach to the steady state is governed by the 
the right eigenvalues 
of the superoperator $\mathcal{S}=\mathcal{H}-\mathcal{M}$, satisfying 
\begin{equation}
\label{eigenvalues}
\mathcal{S}\ket{\Gamma_i}=\lambda_i\ket{\Gamma_i}. 
\end{equation}
The eigenvalues whose real parts are closest to zero thus determine the
asymptotics of the decoherence process. In the following, we refer to
\begin{equation}
  \label{eq:ADR-cm}
  \Delta = \max\left\{
    |\Re\lambda_i|\not=0:\exists\Gamma_i \,\mbox{s.th.}\,
    (\mathcal{S}-\lambda_i)\ket{\Gamma_i}=0 \right\},  
\end{equation}
i.e., the asymptotic decay rate on the level of CMs simply as
ADR.

\subsection{Quadratic and Hermitian Lindblad operators}
\label{quadratic}
The second class of master equations leading to closed equations for the
CM is of the form Eq.~\eqref{Lind1} with Lindblad operators
that are quadratic and Hermitian, as in Eq.~\eqref{definition quadratic}.
Lindblad equations with Hermitian Lindblad operators describe the dynamics of
systems in contact with a classical bath. Let us choose a fluctuating external
field as the source of decoherence (see Sec.~\ref{Experiment}). If,
additionally, the Lindblad operators are quadratic, the fluctuating
Hamiltonian is quadratic. Thus in this case Gaussian states evolve into
mixtures of Gaussian states under such evolutions and we can expect a closed
equation for the CM.

Before discussing the master equation in the CM formalism, let us first
determine in general the steady-state density matrices [see
Eq.~\eqref{definition steadystate}] of a master equation with only Hermitian
Lindblad operators. In that case, we can rewrite the master equation in terms
of $\ket{\rho}$, the vector of components of $\mathbf{\rho}$ as
\begin{equation}
\label{Lindblad3}
\partial_t \ket{\rho}=\bcS\ket\rho=\kla{\bcH-\frac{1}{2}\sum_\alpha\bigl(\bcL^\alpha\bigr)^2}\ket\rho.
\end{equation}
with the superoperators
\begin{align}
\label{superoperator}
\bcH&=-i\kla{\mathbf{H}\otimes\mathbf{1}-\mathbf{1}\otimes\mathbf{H}^\text{T}},\\
\bcL^\alpha&=\mathbf{L}^\alpha\otimes\mathbf{1}-\mathbf{1}\otimes\mathbf{L}^{\alpha\text{T}}.
\end{align}
We observe that the superoperator $\bcH$ is anti-Hermitian and
that the superoperators $\bcL^\alpha$ are Hermitian, so that the
$\bigl(\bcL^\alpha\bigr)^2$ are Hermitian and non-negative.

We consider all complex valued vectors $\ket\rho$ instead of just the ones
corresponding to positive density matrices with trace one. Therefore, we have
to check after the calculation if our results correspond to physically
meaningful states.  The steady states satisfy
\begin{equation}
\label{steadystates}
\bra{\rho_0} \klasmall{\bcH-\frac{1}{2}\sum_\alpha\bigl(\bcL^\alpha\bigr)^2}\ket{\rho_0}=0.
\end{equation}
As stated above, $\bcH$ is anti-Hermitian and all
$\bigl(\bcL^\alpha\bigr)^2$ are Hermitian. Applying these
properties we can conclude from Eq.~\eqref{steadystates} that
\begin{equation}
 \bra{\rho_0}\sum_\alpha\bigl(\bcL^\alpha\bigr)^2\ket{\rho_0}=\bra{\rho_0}\bcH\ket{\rho_0}=0
\end{equation}
holds. It follows from the non-negativity of $\bigl(\bcL^\alpha\bigr)^2$ that
\begin{equation}
 \bigl(\bcL^\alpha\bigr)^2\ket{\rho_0}=0 \hspace{0.2 cm} \forall \alpha.
\end{equation}
Because the $\bcL^\alpha$ can be diagonalized this implies $\bcL^\alpha\ket{\rho_0}=0$. It follows that $\bcH\ket{\rho_0}$ vanishes identically. In terms of matrices $\mathbf{\rho}_0$, we can summarize these conditions for steady states
\begin{equation}
\label{condition_steadystates}
 \left[\mathbf{H},\mathbf{\rho}_0\right]=\left[\mathbf{L^\alpha},\mathbf{\rho}_0\right]=0 \hspace{0.2 cm} \forall \alpha.
\end{equation}
It can be verified with Eq.~\eqref{Lind1} that this condition for steady
states is not only necessary but also sufficient. To summarize, steady states
for Hermitian Lindblad operators correspond to density matrices commuting with
the Hamiltonian and all Lindblad operators. Therefore, they are the identity
up to symmetries shared by the Hamiltonian and the Lindblad operators.

Let us now return to exactly solvable systems in the CM formalism. For
quadratic and Hermitian Lindblad operators and quadratic Hamiltonians the
Master Equation \eqref{Lind1} becomes 
\begin{equation}
\label{Lindblad_covariance}
\partial_t \Gamma=\left[H,\Gamma\right]+\frac{1}{2}\sum_\alpha \left[L^\alpha,\left[L^\alpha,\Gamma\right]\right].
\end{equation}
We can again reformulate this equation for the vector of components $\ket\Gamma$
\begin{equation}
\label{superoperatorcovariance}
\partial_t
\ket{\Gamma}=S\ket\Gamma=\kla{\mathcal{H}-\frac{1}{2}\sum_\alpha\bigl(\mathcal{L}^\alpha\bigr)^2}\ket\Gamma, 
\end{equation}
with $\mathcal{H}$ as in Eq.~(\ref{superoperator_linear}) and
$\mathcal{L}^\alpha= L^\alpha\otimes\id-\id\otimes L^\alpha$.

Since we found that steady states are trivial for Hermitian Lindblad
operators, we concentrate on the asymptotics of the decoherence process. It is
studied through the eigenvalues $\lambda_i$ of the superoperator
$\mathcal{S}$, and in particular its ADR as defined in Eq.~\eqref{eq:ADR-cm}.

\subsection{Translationally invariant Hamiltonians}
\label{sec.tranlationally}
Naturally, translationally invariant systems are best treated in a Fourier
transformed picture. Any real antisymmetric matrix can be transformed into a
real and antisymmetric block-diagonal matrix by an orthogonal transformation
$O$. For the Hamiltonian matrix $H$ this means
\begin{equation}
 H'_{mn,uv}=\kla{OHO^T}_{mn,uv},\hspace{0.1 cm} H'_{mn}=\delta_{mn}\begin{pmatrix} 0&\epsilon_m\\-\epsilon_m&0\end{pmatrix},
\end{equation}
where the real number $\epsilon_m$ are the energies of the elementary
excitations. We, however, transform the Hamiltonian matrix with the unitary
Fourier transform
\begin{equation}
\label{eq.fourier}
 \widetilde{H}_{mn,uv}=\kla{UHU^\dagger}_{mn,uv},\hspace{0.1 cm} \hspace{0.1 cm}
U_{mn,uv}=\frac{1}{\sqrt{N}}e^{\frac{2\pi i}{N}mn}\delta_{uv}.
\end{equation}
The resulting matrix $\widetilde H$ is anti-Hermitian, but not real.  For
translationally invariant systems, for which the $2\times 2$ matrices $H_{jk}$
in Eq.~\eqref{eq:HamMatrix} depend only on $j-k$, the matrix $\widetilde H$ is
block-diagonal with
\begin{equation}
\label{eq.block_diagonal}
\widetilde{H}_{mn}=\delta_{mn}\sum_{s=0}^{N-1}H_s e^{-\frac{2\pi i}{N}s m}.
\end{equation}
The block-diagonal is parameterized according to
\begin{equation}
\label{Hamiltonian_Fourier}
 \widetilde{H}_{nn}=\begin{pmatrix} ik_n & h_n\\-h_n^* & il_n \end{pmatrix}, \hspace{0.2 cm} k_n,l_n\in\mathbb{R},\hspace{0.1 cm}h_n\in\mathbb{C}.
\end{equation}
For later use, we observe the properties
\begin{equation}
\label{symmetries}
 h_{-n}=h_n^*,\hspace{0.2 cm} k_{-n}=-k_n,\hspace{0.2 cm} l_{-n}=-l_n,
\end{equation}
which follow directly from Eq.~\eqref{eq.block_diagonal} for real $H_s$.

For a system that is also invariant under reflections (in real space)
$H_s = -H_s^T$ holds (in addition to $H_{-s} = -H_s^T$ implied by
antisymmetry). In that case, we have $\tilde{H}_{nn} = -\tilde{H}_{nn}$ and
therefore
\begin{equation}
k_n=l_n=0.
\end{equation}

The spectrum of the Hamiltonian matrix determines the elementary
excitation energies
\begin{equation}
\label{elementary excitation energies}
\epsilon_n=\left|\frac{k_n+l_n}{2}\pm\sqrt{\kla{\frac{k_n-l_n}{2}}^2+|h_n|^2}\right|.
\end{equation}
It will be necessary to transform the CM $\Gamma$ accordingly,
defining  
\begin{equation}\label{eq:tildeGamma}
 \widetilde{\Gamma}=U\Gamma U^\dagger.
\end{equation}
By minimizing the energy expectation value
\begin{equation}
 \langle E \rangle={\rm Tr} \klasmall{H^T\Gamma}={\rm Tr} \klasmall{\widetilde{H}^\dagger\widetilde{\Gamma}},
\end{equation}
we find the CM for the ground state. In the case
$k_nl_n<|h_n|^2$ it is
\begin{equation}
\label{ground state covariance matrix}
 \widetilde{\Gamma}^0_{mn}=\delta_{mn}{\scriptstyle \left[\klasmall{\frac{k_n-l_n}{2}}^2+|h_n|^2\right]^{-1/2}}
 \begin{pmatrix} i\frac{k_n-l_n}{2}&-h_n\\h_n^*&-i\frac{k_n-l_n}{2}\end{pmatrix}
\end{equation}
and otherwise
\begin{equation}
\label{eq.isotropic}
 \widetilde{\Gamma}^0_{mn}=-i\delta_{mn}{\rm sign}\kla{k_n+l_n}
 \id_2.
\end{equation}
For translationally invariant and reflection symmetric systems $k_nl_n=0$
holds, thus $k_nl_n<|h_n|^2$ is fulfilled in such systems. Since the XY chain
Eq.~\ref{XY chain} is reflection symmetric, we can concentrate on the case of
Eq.~\eqref{ground state covariance matrix}. Specifically, we obtain for the
Hamiltonian Eq.~\ref{XY chain} that
\begin{gather}
\label{hn chain}
 h_n=-2B+2J\left[(1+\gamma)e^{\frac{2\pi i}{N}n}+(1-\gamma)e^{-\frac{2\pi
       i}{N}n}\right],\\
\hspace{0.2 cm} k_n=l_n=0, 
\end{gather}
which contains a continuous quantum phase transition at $B=2J$, where the gap
closes and an elementary excitation energy $\epsilon_n=|h_n|=0$ exists.  This
Hamiltonian will be further discussed in Sec.~\ref{section XY chain}.

\section{Linear Lindblad operators}
\label{sec:applic:linear}

Now we apply the formalism introduced in the previous Sections to some simple
cases of physical interest.  Here we choose the simplest examples, i.e.,
linear Lindblad operators (see Sec.~\ref{linear}).  We study two settings. In
Sec.~\ref{linear no Hamiltonian} we look at systems without any unitary
evolution, observing dynamic transitions when tuning the strength of competing
decoherence processes. Here we enrich our presentation with an example for
dissipative state engineering. In Sec.~\ref{linear with Hamiltonian} we
consider open systems governed by a Hamiltonian, which describes a quantum
phase transition itself, and show that the dissipative system undergoes a
transition for the same values of the system parameters.

\subsection{Purely dissipative systems}
\label{linear no Hamiltonian}
The simplest example of two competing decoherence processes generated by
linear Lindblad operators is
\begin{gather}
\label{Lindblad local}
\mathbf{L}^\alpha_-=g\mu a_\alpha, \hspace{0.5cm}
\mathbf{L}^\alpha_+=g\nu a^\dagger_\alpha,
\end{gather}
acting on site $\alpha \in\{1,\dots,N\}$. It describes the competition between
particle-loss and particle-gain processes. We observe that the Master
Equation~\eqref{Lindblad_linear} without the Hamiltonian ($H=0$) is diagonal
in real space
\begin{multline}
 \partial_t\Gamma=-g^2(\mu^2+\nu^2)\Gamma
-g^2(\mu^2-\nu^2)\bigoplus_{\alpha=1}^N (i\sigma_y).
\end{multline}
In this simple case the master equation is already diagonal and we read off 
the single decoherence rate
$\Delta=g^2(\mu^2+\nu^2)$. Solving the master equation for
$\partial_t\Gamma_0=0$ gives the unique steady-state CM
\begin{equation}\label{eq:GammassLinL}
 \Gamma_0=-\frac{\mu^2-\nu^2}{\mu^2+\nu^2}\bigoplus_{\alpha=1}^N\begin{pmatrix}0&1\\-1&0\end{pmatrix},
\end{equation}
which is block diagonal. This state is characterized by the particle number
$\langle a_\alpha^\dagger a_\alpha\rangle=\nu^2/(\mu^2+\nu^2)$ at all
sites. For pure particle-loss processes $(\nu=0)$, all sites are unoccupied
$\langle a_\alpha^\dagger a_\alpha\rangle=0$ in the steady state, while for pure
particle-gain processes $(\mu=0)$, all sites are occupied $\langle
a_\alpha^\dagger a_\alpha\rangle=1$. At $\mu=\nu$ the steady state is the
unpolarized completely mixed state.
Not surprisingly, the system does not display any phase
transition. 

More interesting may be the case in which dissipation can also induce
correlations. A simple example of this kind is provided by the Lindblad
operators 
\begin{equation}
\label{Lindblad paired}
 \mathbf{L}^\alpha=g\kla{\mu a_\alpha+ \nu a^\dagger_{\alpha+1}}
\end{equation}
acting on nearest neighbors. This set of Lindblad operators generates a master
equation, which is diagonal after the Fourier transform \eqref{eq.fourier}
\begin{align}
 \partial_t\widetilde{\Gamma}=&
-g^2(\mu^2+\nu^2)\widetilde{\Gamma}\\
&-g^2\mu\nu\left\{\bigoplus_{n=1}^N\cos(2\pi n/N)\sigma_z,\widetilde{\Gamma}\right\} \nonumber\\
&-g^2(\mu^2-\nu^2)\bigoplus_{n=1}^Ni\sigma_y\nonumber\\
&-2g^2\mu\nu\bigoplus_{n=1}^N i\sin(2\pi n/N)\sigma_x.\nonumber
\end{align}
In this case, a spectrum of decoherence rates
$g^2\{\mu^2+\nu^2\pm2\mu\nu[\cos\frac{2\pi n}{N}+\cos\frac{2\pi m}{N}],
\mu^2+\nu^2\pm2\mu\nu[\cos\frac{2\pi n}{N}-\cos\frac{2\pi m}{N}]\}$
exists with a ``gap'' $g^2(\mu-\nu)^2$. The unique steady state is 
\begin{multline}
\label{covariance paired}
\widetilde\Gamma_0=-\frac{\mu^2-\nu^2}{\mu^2+\nu^2}\bigoplus_{n=1}^Ni\sigma_y-\frac{2\mu\nu}{\mu^2+\nu^2}\bigoplus_{n=1}^Ni\sin(2\pi n/N)\sigma_x.
\end{multline}
This state is a paired fermionic state according to the definition of Kraus
\emph{et al.} \cite{KWCG08}. Paired states show two-particle quantum
correlations that can not be be reproduced by separable states (mixtures of
Slater determinants). It is proven in \cite{KWCG08} that Gaussian states are
paired iff $Q_{kl}=\langle \frac{i}{2}[a_k,a_l]\rangle\ne 0$. This condition
expresses the fact that separable states are convex combinations of states
with a fixed particle number. For the CM \eqref{covariance
  paired} we get
\begin{equation}
Q_{kl}=\begin{cases}
	  \frac{1}{2}\frac{\mu\nu\cdot \text{sign}\kla{k-l}}{\mu^2+\nu^2} & \text{if } |k-l|=1\\
	  0 & \text{if } |k-l|\ne 1
       \end{cases}.
\end{equation}
We conclude that \eqref{Lindblad paired} generates paired states, except for
the trivial cases $\mu=0$ or $\nu=0$. Note that even though the gap closes at
$\mu=\nu$ (where maximal pairing is created) there is no phase transition at
this point. 

\subsection{Dissipative systems with Hamiltonians}
\label{linear with Hamiltonian}
A different form of transitions can arise in the presence of
a Hamiltonian when tuning the parameters of the Hamiltonian. To show this, we
solve the evolution of the Lindblad master equation \eqref{Lindblad_linear}
with a general quadratic and translationally invariant Hamiltonian [see
Eqs.~\eqref{eq:HamMatrix} and \eqref{Hamiltonian_Fourier}]. We choose the local
Lindblad operators \eqref{Lindblad local}, again because they are the simplest
example. The diagonal master equation in Fourier space becomes
\begin{align}\label{eq:LinLb+HamMEq}
  \partial_t\widetilde\Gamma=&\left[\bigoplus_{n=1}^N\begin{pmatrix} ik_n & h_n\\-h_n^* & il_n \end{pmatrix},\widetilde\Gamma\right] \nonumber\\
&-g^2(\mu^2+\nu^2)\widetilde\Gamma-g^2(\mu^2-\nu^2)\bigoplus_{n=1}^Ni\sigma_y.
\end{align}
The corresponding steady-state CM in the weak-coupling limit
$g\rightarrow 0$ is \footnote{The exact steady state CM for
  Eq.~(\ref{eq:LinLb+HamMEq}) differs from $\tilde\Gamma_0$ of
  Eq.~(\ref{eq:LinLb+HamSS}) by
  $i2g^2\bigoplus_n\frac{\mu^2-\nu^2}{(k_n-l_n)^2+4|h_n|^2}\left(-\Im(h_n)\sigma^z
    + \frac{k_n-l_n}{2}\sigma^x\right)$ which vanishes as $g\to0$.}
\begin{multline}\label{eq:LinLb+HamSS}
 \widetilde\Gamma_0=-\frac{\mu^2-\nu^2}{\mu^2+\nu^2}\bigoplus_{n=1}^N\frac{\Re\kla{h_n}}{(k_n-l_n)^2/4+|h_n|^2}\\\cdot\begin{pmatrix}
				   i(k_n-l_n)/2 & h_n\\
				   -h^*_n & -i(k_n-l_n)/2 
			        \end{pmatrix}.
\end{multline}
Transforming back to $\Gamma_0$ [and using Eq.~\eqref{symmetries}] we can read
off the particle number $\langle 2a_j^\dagger a_j-1\rangle=
(\Gamma_0)_{jj,01}$ as
\begin{equation}
\label{steady state linear example}
 (\Gamma_0)_{jj,01}=\frac{1}{2}\frac{\mu^2-\nu^2}{\mu^2+\nu^2}\frac{1}{N}\sum_{n=1}^N\frac{\Re\kla{h_n}^2}{(k_n-l_n)^2/4+|h_n|^2}.
\end{equation}
Based on this result we can now discuss how non-analytic behavior in the
steady state correlates with critical points of the system. A vanishing
denominator in Eq.~\eqref{steady state linear example} is not a priori a
sufficient condition for non-analytic behavior because the numerator might
vanish at the same point. This is relevant for interesting examples with
$k_n-l_n=0$, e.g., the XY chain in Eq.~\eqref{hn chain}. We give a rigorous
discussion in the following. In the thermodynamic limit, the sums over
expectation values in Eq.~\eqref{steady state linear example} can be replaced
by a loop integral around the origin of the complex plane with radius one,
where the integration variable is $z=\exp\klasmall{\frac{2\pi i}{N}n}$. This
is possible because $h_n$, $k_n$, and $l_n$ are Fourier series. For local
interactions, the denominator of the integrand is a polynomial in $z$ [see
Eq.~\eqref{eq.fourier}] and thus has a finite number of distinct
roots. Applying the residue theorem, a non-analyticity in $\langle a_j^\dagger
a_j\rangle$ is possible only if a residue of the integrand, i.e., a root of
its denominator, moves through the integral contour in the complex plane as a
function of some external parameters. This happens for a vanishing denominator
$|h_n|^2+(k_n-l_n)^2/4=0$ for some real $n\in [0,N)$. In the special case of a
reflection symmetric system $k_n+l_n=0$ this coincides with a vanishing energy
gap $\epsilon_n=0$ [see Eq.~\eqref{elementary excitation energies}], a
signature for a quantum phase transition. To summarize, for a reflection
symmetric system with $|h_n|^2+(k_n-l_n)^2/4=0$ in the weak-coupling limit a
quantum phase transition occurs in the dissipative system for the same
parameter values as in the corresponding closed system and is signaled by a
non-analyticity in $\langle a_j^\dagger a_j\rangle$. This calculation is
explicitly performed in section \ref{linear XY chain} for the XY chain
\footnote{The ADR itself $g^2(\mu^2+\nu^2)$ does not change at the phase
  transition. However, note that there is actually a large manifold of
  eigenvalues of the Liouvillian whose real part is $-g^2(\mu^2+\nu^2)$ but
  with different imaginary parts, taken from the set
  $\{\lambda^{(n)}_\pm-\lambda^{(m)}_\pm\}$, where $\lambda^{(n)}_\pm =
  (k_n+l_n)/2\pm \sqrt{(\frac{k_n-l_n}{2})^2+|h_n|^2}$ and all four
  combinations of the subscripts $\pm$ may occur.  At the critical point the
  discriminant vanishes and each eigenvalue becomes (at least) fourfold
  degenerate.}.

\section{Quadratic and Hermitian Lindblad operators}
\label{sec:applic:quadratic}
In this Section we turn to the dynamical properties of the Lindblad master
equation with quadratic and Hermitian Lindblad operators as introduced in
Sec.~\ref{quadratic}. In the study of closed systems, quantum phase
transitions are signaled by non-analyticities in ground state expectation
values. In the dissipative case the steady state is the analog of the ground
state. However, we have shown in Sec.~\ref{quadratic} that in the case of
Hermitian Lindblad operators the steady states are trivial and thus cannot
evidence a phase transition. Therefore we turn to the ADR, which determines
the long-time dynamics of the decoherence process. We identify non-analytical
behavior of this rate both in the absence of any Hamiltonian (see
Sec.~\ref{quadratic no Hamiltonian}) for competing decoherence processes and
for non-zero Hamiltonian, in which case phase transitions of the corresponding
closed system are reflected in a ``dynamical transition'' of this 
rate (see Sec.~\ref{quadratic with Hamiltonian}).

\subsection{Purely dissipative systems}
\label{quadratic no Hamiltonian}
A particular simple set of local and quadratic Lindblad operators is
\begin{gather}
\label{Lindblad operators quadratic1}
\mathbf{L}^\alpha_z=g\mu\frac{i}{2}\left[c_{\alpha,1},c_{\alpha,0}\right],\\
\label{Lindblad operators quadratic2}
\mathbf{L}^\alpha_x=g\nu\frac{i}{2}\left[c_{\alpha+1,0},c_{\alpha,1}\right].
\end{gather}
In this case the Lindblad equation \eqref{Lindblad_covariance} becomes
\begin{align}
\label{Lindbladsigmaz}
 \partial_t \Gamma_{kl,uv}=&-4g^2\mu^2\Gamma_{kl,uv}\kla{1-\delta_{kl}}\\
&-4g^2\nu^2\Gamma_{kl,uv}(1-\delta_{2k+u+1,2l+v}\delta_{k+1,l}\nonumber\\
&\hspace{2.6cm}-\delta_{2k+u-1,2l+v}\delta_{k-1,l}),\nonumber
\end{align}
We can read off the decoherence rates $-4g^2(\mu^2+\nu^2)$, $-4g^2\mu^2$, and
$-4g^2\nu^2$. Thus, the ADR
\begin{equation}
 \Delta=\begin{cases}
      4g^2\mu^2 & \text{if } \mu\le\nu\\
       4g^2\nu^2 & \text{if } \nu<\mu
     \end{cases}
\end{equation}
undergoes a dynamical transition as a function of $\mu/\nu$ at $\mu=\nu$.

\subsection{Dissipative systems with Hamiltonian}
\label{quadratic with Hamiltonian}
Now we add a quadratic Hamiltonian and calculate the ADR $\Delta$ in the limit of small couplings to the environment $g\rightarrow
0$. First, we derive it for the quadratic Lindblad operators from
Eqs.~(\ref{Lindblad operators quadratic1},\ref{Lindblad operators quadratic2})
for $\nu=0$ and $\mu=1$. Later we 
will present the results for the case of arbitrary $\mu$ and $\nu$. For
translationally invariant systems the Fourier transformed master equation
\eqref{Lindbladsigmaz} is 
\begin{multline}
\label{LindbladsigmazFourier}
\partial_t \widetilde{\Gamma}_{kl}\equiv(\widetilde{\mathcal{S}}\widetilde{\Gamma})_{kl}=\bigl[\widetilde{H},\widetilde{\Gamma}\bigr]_{kl}\\-4g^2\kla{\widetilde{\Gamma}_{kl}-\frac{1}{N}\sum_{r,s=1}^{N}\widetilde{\Gamma}_{rs}\delta_{r-s,k-l}},
\end{multline}
with the 
unitarily transformed superoperator
$\widetilde{\mathcal{S}}$ to 
\begin{equation}
 \widetilde{\mathcal{S}}=\klasmall{U\otimes U}\mathcal{S}\klasmall{U\otimes U}^\dagger,
\end{equation}
with $U$ from Eq.~\eqref{eq.fourier}. For weak couplings between system and
bath $g\rightarrow 0$, the eigenvalues of $\widetilde{\mathcal{S}}$ (and thus
of $\mathcal{S}$) can be determined by first order perturbation expansion. To
this end we first diagonalize the unperturbed Hamiltonian part of
$\widetilde{\mathcal{S}}$
\begin{equation}
 \bigl[\widetilde{H},\widetilde{\Gamma}\bigr]_{kl}=\widetilde{H}_{kk}\widetilde{\Gamma}_{kl}-\widetilde{\Gamma}_{kl}\widetilde{H}_{ll}\stackrel{!}{=}\lambda\widetilde{\Gamma}_{kl},
\end{equation}
where we use the notation introduced in Eq.~\eqref{eq.block_diagonal} for the
Hamiltonian $\widetilde{H}$. The $4N^2$ eigenvalues $\lambda^{mna}$
($m,n=1,\dots, N$, $a=1,\dots, 4$) are
\begin{gather}
 \lambda^{mn1}=i\kla{\alpha_m-\alpha_n+\beta_m-\beta_n},\\
 \lambda^{mn2}=i\kla{\alpha_m-\alpha_n-\beta_m+\beta_n},\\
 \lambda^{mn3}=i\kla{\alpha_m-\alpha_n+\beta_m+\beta_n},\\
 \lambda^{mn4}=i\kla{\alpha_m-\alpha_n-\beta_m-\beta_n},
\end{gather}
with
\begin{gather}
 \alpha_m=|k_m+l_m|/2,\\
 \beta_m=\sqrt{|h_m|^2+(k_m-l_m)^2/4}.
\end{gather}
The corresponding eigenmatrices are denoted as $\widetilde{\Lambda}^{mna}$
with nonzero elements $\widetilde{\Lambda}^{mna}_{kl}$ only for $m=k$ and
$n=l$, i.e.,
$\widetilde{\Lambda}^{mna}_{kl}=\delta_{mk}\delta_{nl}\widetilde{\Lambda}^{mna}_{mn}$.
Perturbation theory demands to calculate the matrix elements of the
perturbative part of $\widetilde{\mathcal{S}}$,
$-4g^2(\delta_{mk}\delta_{nl}\delta_{ab}-P_{klb}^{mna}/N)$ [see
Eq.~\eqref{LindbladsigmazFourier}], with
\begin{eqnarray}
P^{mna}_{klb}&=&\frac{N}{4g^2}\bra{\widetilde{\Lambda}^{mna}}\frac{1}{2}\sum_\alpha\klasmall{\mathcal{L}^\alpha}^2\ket{\widetilde{\Lambda}^{klb}}+N\delta_{mk}\delta_{nl}\delta_{ab},\nonumber\\
&=&\sum_{q,r,s,t=1}^{N}\delta_{q-r,s-t}{\rm Tr}\left[\klasmall{\widetilde{\Lambda}^{mna}_{st}}^\dagger\widetilde{\Lambda}^{klb}_{qr}\right],\nonumber\\
&=&\delta_{m-n,k-l}{\rm Tr}\left[\klasmall{\widetilde{\Lambda}^{mna}_{mn}}^\dagger\widetilde{\Lambda}^{klb}_{kl}\right].
\end{eqnarray}
Thus the eigenvalues of $\widetilde{\mathcal{S}}$ are determined by those of the
Hermitian matrix $P$ and the largest eigenvalue smaller than $N$ of 
$P$ (restricted to a space of degenerate eigenvalues $\lambda^{mna}$ of
$[\tilde{H},\cdot]$) determines the ADR. We denote it by $\Delta_P$ and 
thus have that the ADR is
$\Delta=4g^2\klasmall{1-\frac{\Delta_P}{N}}$.  
To find $\Delta_P$, note that the matrix elements of $P$ fulfill
$\bigl|P_{klb}^{mna}\bigr|\le 1$. Thus an $N$-fold degeneracy of
$\lambda^{mna}$ is required for $\Delta_P=\Omega(N)$. Generically, this is
possible only for the eigenvalue $\lambda^{mna}=0$, i.e., $m=n$ and
$a=1,2$. The corresponding eigenmatrices are
\begin{align}
\widetilde{\Lambda}^{mm1}_{kl}&=\delta_{mk}\delta_{ml}\frac{1}{\sqrt{2}\beta_n}\begin{pmatrix} i\frac{k_m-l_m}{2}&-h_m\\h_m^*&-i\frac{k_m-l_m}{2}\end{pmatrix},\\
\widetilde{\Lambda}^{mm2}_{kl}&=\delta_{mk}\delta_{ml}\frac{1}{\sqrt{2}}\begin{pmatrix} 1&0\\0&1\end{pmatrix}.
\end{align}
As the eigenmatrices $\widetilde{\Lambda}^{mm2}$ give eigenvalues equal to $N$
and $0$ only, have no overlap with physical CMs, and yield
$P_{kl1}^{mn2}=0$, we focus on the matrices $\widetilde{\Lambda}^{mm1}$. The
corresponding part of the perturbation matrix is
\begin{equation}
 P_{mn}=P^{mm1}_{nn1}
=\frac{2h_mh_n^*+2h_m^*h_n-\klasmall{k_m-l_m}\klasmall{k_n-l_n}}{4\beta_m\beta_n}.
\end{equation}
We diagonalize this matrix by introducing the three vectors $\ket{a},\ket{b},\ket{c}\in\mathbb{C}^N$ with the components
\begin{equation}
\label{three vectors}
 a_m=\frac{k_m-l_m}{2\beta_m},\hspace{0.1 cm} b_m=\frac{\Im\klasmall{h_m}}{\beta_m},\hspace{0.1 cm} c_m=\frac{\Re\klasmall{h_m}}{\beta_m},
\end{equation}
and writing $P_{mn}$ in terms of these unnormalized vectors
\begin{equation}
 P=\ket{c}\bra{c}+\ket{b}\bra{b}-\ket{a}\bra{a}.
\end{equation}
We now exploit the symmetries of $h_n$, $k_n$, and $l_n$ stated in
Eq.~\eqref{symmetries}. First we observe that $\ket{c}$ is orthogonal to
$\ket{a}$ and $\ket{b}$. We have chosen the CMs corresponding
to the three vectors \eqref{three vectors} anti-Hermitian, since this matrix
remains anti-Hermitian even in the complex vector space. After
transforming back into real space the ones corresponding to $\ket{a}$ and
$\ket{b}$ are purely imaginary so that they have no overlap with any
physically meaningful real and antisymmetric CM. Only the
matrix corresponding to $\ket{c}$ is real and antisymmetric and given by 
\begin{equation}
 \Gamma_\Delta=\kla{\sum_{m=1}^N\frac{|h_m|^2}{2\beta_m^2}}^{-1}\sum_n\frac{\Re\klasmall{h_n}}{\beta_n}U^\dagger\Lambda^{nn1}U.
\end{equation}
Therefore, it determines the ADR. We get
\begin{equation}
 \Delta_P=\sum_{m=1}^N \frac{\Re\klasmall{h_m}^2}{\beta_m^2}=\sum_{m=1}^N \frac{\Re\klasmall{h_m}^2}{|h_m|^2+(k_m-l_m)^2/4},
\end{equation}
and thus 
\begin{equation}
\label{eq:example ADR}
 \Delta=\frac{4g^2}{N}\sum_{m=1}^N \frac{4\Im\klasmall{h_m}^2+(k_m-l_m)^2}{4|h_m|^2+(k_m-l_m)^2}
\end{equation}
as the general from of the ADR. 

We can extend our analysis to systems with the general Lindblad operators
Eqs.~(\ref{Lindblad operators quadratic1},\ref{Lindblad operators quadratic2})
and find in an analog way the two lowest decay rates
\begin{equation}
\label{full decoherence rates}
 \frac{\Delta_\pm}{4g^2}=\mu^2+\nu^2-\frac{\epsilon_z+\epsilon_x}{2}\pm\sqrt{\kla{\frac{\epsilon_z-\epsilon_x}{2}}^2+\epsilon^2},
\end{equation}
with
\begin{align}
\label{sums}
\epsilon_z&=\frac{\mu^2}{N}\sum_{m=1}^N \frac{\Re\klasmall{h_m}^2}{\beta_m^2},\nonumber\\
\epsilon_x&=\frac{\nu^2}{N}\sum_{m=1}^N \frac{\Re\klasmall{h_m\exp(-2\pi im/N)}^2}{\beta_m^2},\nonumber\\
\epsilon&=\frac{\mu\nu}{N}\sum_{m=1}^N \frac{\Re\klasmall{h_m}\Re\klasmall{h_m\exp(-2\pi im/N)}}{\beta_m^2}.
\end{align}
We can now argue that the ADR itself reflects the criticality of the system.
The argument is completely analogous to the one given in Sec.~\ref{linear with
  Hamiltonian}. If the denominator becomes zero, we can expect a
non-analyticity expressions Eqs.~(\ref{sums}). In particular, in the
reflection symmetric case $k_n+l_n=0$, where the denominator agrees with the
elementary excitation energies [$\epsilon_n^2=|h_n|^2+(k_n-l_n)^2/4=0$, see
Eq.~\eqref{elementary excitation energies}] the non-analyticity in the ADR
signals the presence of a quantum phase transition in the Hamiltonian itself.

\section{Example Hamiltonians}
\label{section XY chain}
In this section we will revisit the results obtained for the steady state and
the ADR for linear and quadratic Lindblad operators in Secs.~\ref{linear with
  Hamiltonian} and \ref{quadratic with Hamiltonian} for the specific
Hamiltonian \eqref{XY chain} of the quantum XY chain.

The energies of the elementary excitations of this Hamiltonian are
$\epsilon_n=|h_n|$. Thus, for the XY chain in Eq.~\eqref{hn chain} the gap
closes at $B=2J$ in the thermodynamic limit and the quantum XY chains exhibit
a phase transition at this point. In fact, these models constitute the
archetypal example of a continuous quantum phase transition \cite{Sachdev}. In
this chapter we want to find properties of the dissipative dynamics signaling
this phase transition.

\subsection{Linear Lindblad operators}
\label{linear XY chain}
Let us now apply the findings from Sec.~\ref{linear with Hamiltonian} and Eq. ~\eqref{steady state linear example} to the example system defined in Eq.~\eqref{hn chain} which contains a quantum phase transition at $B=2J$. Then the particle numbers become for $k_n=l_n=0$
\begin{equation}
\label{particle number linear Lindblad example}
 \langle2a_n^\dagger a_n-1\rangle=\frac{1}{2}\frac{\mu^2-\nu^2}{\mu^2+\nu^2}\kla{1+\frac{1}{N}\sum_{n=1}^N\frac{h_n^*}{h_n}}.
\end{equation}
For $\gamma=0$ we easily obtain $\Delta=0$ (since by Eq.~\eqref{hn chain}
$h_n$ is real in that case and then by Eq.~\eqref{eq:example ADR} $\Delta$ is
zero for $k_n=l_n=0$ ). For $\gamma\ne 0$ we evaluate the sum
$1/N\cdot\sum_{m=1}^N h_m^*/h_m$ in the thermodynamic limit by introducing the
complex variable $z=\exp\klasmall{-2\pi im/N}$
\begin{multline}\label{eq:contour}
  \lim_{N\rightarrow\infty}\frac{1}{N}\sum_{m=1}^N \frac{h_m^*}{h_m}=\\
  \frac{1}{2\pi
    i}\oint_{|z|=1}\frac{dz}{z}\frac{2J\klasmall{1-\gamma}z^2-2Bz+2J\klasmall{1+\gamma}}{2J\klasmall{1+\gamma}z^2-2Bz+2J\klasmall{1-\gamma}},
\end{multline}
where the integration contour is a circle of radius $|z|=1$ around $z=0$ in the complex plane. The complex integrand is analytic except for three distinct poles at
\begin{align}
 z^0&=0,\nonumber\\
 \label{eq:poles}z^\pm&=\frac{1}{2J\klasmall{1+\gamma}} \left[B\pm\sqrt{B^2-4J^2\klasmall{1-\gamma^2}}\right].
\end{align}
\begin{figure}[h]
  \centering
  \includegraphics[width=0.4\textwidth]{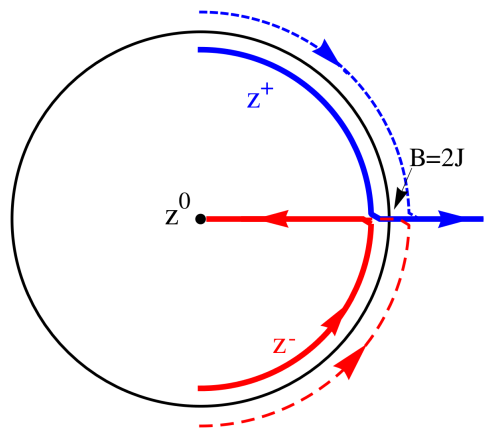}
  \caption{(Color online) The poles $z^0,z^\pm$ [see Eq.~\eqref{eq:poles}] are plotted for
    $J=1,\gamma=\pm0.1$. As $B$ is changed from $0$ to $20$ the poles
    $z^+(z^-)$ for positive anisotropy $\gamma=+0.1$ move along the blue (red)
    solid curves and $z^+$ crosses the contour at the critical value
    $B=2J$. For negative $\gamma=-0.1$, $z^-$ crosses at $B=2J$. At the
    crossing the integral Eq.~\eqref{eq:contour} changes non-analytically.}
  \label{fig:residuum}
\end{figure}
The contour integral is determined by the sum over the residues at those poles
which are inside the contour ($|z|<1$). $z^0$ is always inside this
contour. In the case $\gamma>0$, $z^+$ is inside the contour for $0\leq B<2J$
and outside for $B>2J$, while $z^-$ is always inside the contour. In the case
$\gamma<0$, $z^-$ is inside the contour for $B>2J$ and outside for $0\leq
B<2J$, while $z^+$ is always outside the contour. So residues cross the
contour at the quantum phase transition $B=2J$ (because then $h_n=0$ for some
$n$), leading to a non-analytical behavior in the particle density of the
steady state.

After applying the residue theorem we get the particle number of the steady
state
\begin{multline}
\label{final result linear}
 \langle 2a_n^\dagger a_n-1\rangle=\\
 \frac{\mu^2-\nu^2}{\mu^2+\nu^2}\cdot\begin{cases}
                \frac{1}{1+|\gamma|} & B\le 2J\\
		\frac{1}{1-\gamma^2}\kla{1-\frac{\gamma^2}{\sqrt{1-\kla{\frac{2J}{B}}^2\kla{1-\gamma^2}}}} & B\ge 2J
               \end{cases}
\end{multline}
for all $\gamma$, which does not depend on the sign of $\gamma$. For $B<2J$
the particle number in the steady state does not vary with the magnetic field,
while its magnitude approaches $(\mu^2-\nu^2)/(\mu^2+\nu^2)$ for large
magnetic fields like $\sim \klasmall{J/B}^2$. To summarize, the steady state
undergoes a dissipative phase transition at $B=2J$ signaling the phase
transition in the system.

\subsection{Quadratic and Hermitian Lindblad operators}
\label{sec.example}
\begin{figure}[tb]
\centering
\includegraphics[width=85mm]{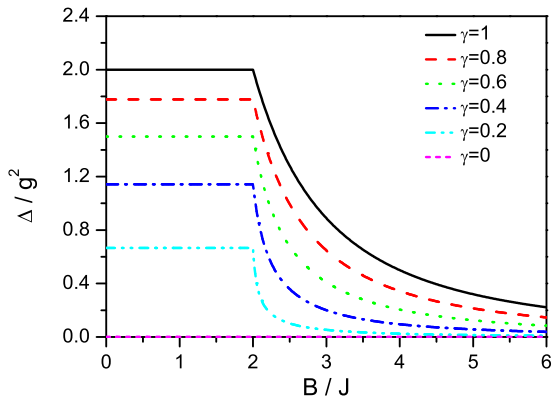}
\caption{(Color online) ADR $\Delta$ [see Eq.~\eqref{eq:example ADR}] of
  the XY chain \eqref{XY chain} for different anisotropy parameters $\gamma$
  as a function of the magnetic field in the limits $N\rightarrow\infty$ and
  $g\rightarrow0$. A phase transition in $\Delta$ is visible at $B=2J$ for
  $\gamma\ne 0$.}
\label{exact mu}
\end{figure}

As an example we study the anisotropic XY chain in a transverse magnetic field
with the Hamiltonian given in Eq.~\eqref{XY chain}. 
This translationally invariant Hamiltonian is
Jordan-Wigner transformed to a quadratic fermionic Hamiltonian with
Hamiltonian matrix $H$ given by 
\begin{gather}\label{eq:XY-fermionic}
 H_0=\begin{pmatrix} 
      0&-2B\\2B&0
     \end{pmatrix},\\
H_1=\begin{pmatrix} 
      0&2J\klasmall{1-\gamma}\\-2J\klasmall{1+\gamma}&0
     \end{pmatrix},\\
H_{-1}=\begin{pmatrix} 
      0&2J\klasmall{1+\gamma}\\-2J\klasmall{1-\gamma}&0
     \end{pmatrix}.
\end{gather}
After Fourier transforming [see Eq.~\eqref{eq.fourier}] this Hamiltonian
matrix assumes the form given in Eq.~\eqref{Hamiltonian_Fourier} with
parameters $h_n, k_n, l_n$ given by \eqref{hn chain}.

We now apply the results from Sec.~\ref{quadratic with Hamiltonian}  
to the Hamiltonian Eq.~\eqref{eq:XY-fermionic} and the Lindblad operators
\begin{equation}
  \mathbf{L}^\alpha= g\mu\frac{i}{2}\left[c_{\alpha,1},c_{\alpha,0}\right]
  \leftrightarrow  g\sigma^\alpha_z.   
\end{equation}
After a brief discussion of the steady states and a derivation of the ADR in
the thermodynamic ($N\rightarrow\infty$) and in weak coupling ($g\rightarrow
0$) limits, we present numerical results of the system dynamics for finite $N$
and $g$ and compare them with our analytic predictions.

First, we discuss the steady states of these systems (see
Sec.~\ref{quadratic}). From Eq.~\eqref{condition_steadystates} we have
concluded that the steady-state density matrix is the identity up to
symmetries shared by the Lindblad operators and the Hamiltonian. A rigorous
derivation of the steady states for this example could start from the ansatz
that the steady-state density matrix is diagonal in the Fock basis, following
from $\bigl[\sigma_z^\alpha,\rho\bigr]=0$. Then the commutator
$\bigl[\mathbf{H},\rho\bigr]=0$ must be exploited to get the steady state.

As the Lindblad operators correspond to local particle number operators, the
important compatible symmetries for the XY chains are the parity
$\mathcal{P}=\sigma_z^1\dots\sigma_z^N$, discriminating between an odd and an
even number of particles, and the total particle number
$\mathcal{N}=\klasmall{\mathbf{1}+\sum\sigma_z^j}/2$. For truly asymmetric XY
chains $\gamma\ne 0.5$ the parity is the highest symmetry compatible with the
Lindblad operators. In these cases the steady-state density matrix is given by
the identity in the two sectors of even and odd parity, the relative weight of
these sectors is determined by the initial state. For the symmetric chain
$\gamma=0.5$, the steady-state density matrix is the identity only in the
sectors with a constant total number of particles. Thus for $\gamma\ne 0.5$
the steady-state magnetization is $\langle\sigma_z^j\rangle=0$ regardless of
the initial state, whereas the magnetization of the initial state is conserved
for $\gamma=0.5$.

Second, we calculate the ADR \eqref{eq:example ADR} for the XY chains with
Eq.~\eqref{hn chain} analogous to the integration in Sec.~\ref{linear XY
  chain}.

\begin{figure}[tb]
\centering
\includegraphics[width=85mm]{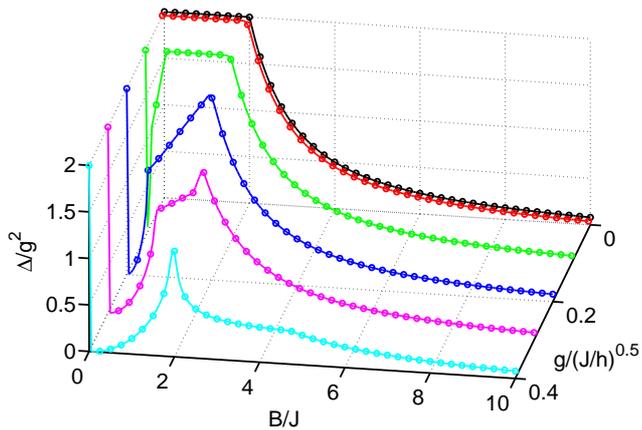}
\caption{(Color online) ADR $\Delta$ [see Eq.~\eqref{eq:example ADR}]
  of the XY chain \eqref{XY chain} for different coupling strengths $g$,
  $\gamma=1$, and $N=100$ as a function of the magnetic field $B$. For $g\le
  0.1\kla{J/\hbar}^{0.5}$ the results agree with the limit of weak coupling
  $g\rightarrow 0$ [see Eq.~\eqref{final result}].}
\label{finite coupling}
\end{figure}

\begin{figure}[tb]
\centering
\includegraphics[width=85mm]{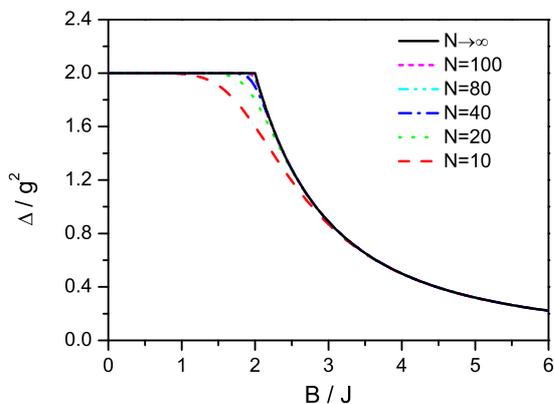}
\caption{(Color online) ADR $\Delta$ [see Eq.~\eqref{final result}]
 of the XY chain \eqref{XY chain} for different system sizes
  $N$ and $\gamma=1$, $g=0.01\kla{J/\hbar}^{0.5}$ as a function of the
  magnetic field $B$. For $N\ge 50$ the thermodynamic limit is reached except
  for small variations at the phase transition $B=2J$.}
\label{finite size}
\end{figure}

After applying the residue theorem we get the ADR
\begin{equation}
\label{final result}
\Delta=4g^2
\begin{cases}
  \frac{|\gamma|}{1+|\gamma|} & B\le 2J\\
  \frac{\gamma^2}{1-\gamma^2}\kla{\left[1-\kla{\frac{2J}{B}}^2\kla{1-\gamma^2}\right]^{-1/2}-1} & B\ge 2J
               \end{cases}
\end{equation}
for all $\gamma$ in the case $\mu=1$ (and $\nu=0$). It does not depend on the
sign of $\gamma$ and is shown in Fig.~\ref{exact mu} for several values of
$\gamma\in[0,1]$. For $B<2J$ the ADR does not vary with the magnetic field,
while for large magnetic fields its magnitude decreases to zero and scales as $(J/B)^2$. The same behavior was found for the variance of the
particle number in these models in a previous work \cite{Sybille}. To
summarize, the ADR undergoes a dissipative phase transition at $B=2J$
signaling the phase transition in the system.

\begin{figure}[tb]
\centering
\includegraphics[width=85mm]{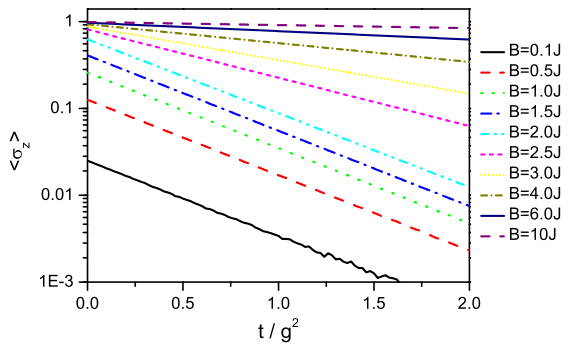}
\caption{(Color online) Evolution of the magnetization $\langle \sigma^j_z\rangle$ in time
  starting from the system ground state of the XY chain \eqref{XY chain}, for
  different magnetic fields $B$, $g=0.01\kla{J/\hbar}^{0.5}$, and
  $\gamma=1$. The magnetization decreases exponentially in time.}
\label{dynamics}
\end{figure}

\begin{figure}[tb]
\centering
\includegraphics[width=85mm]{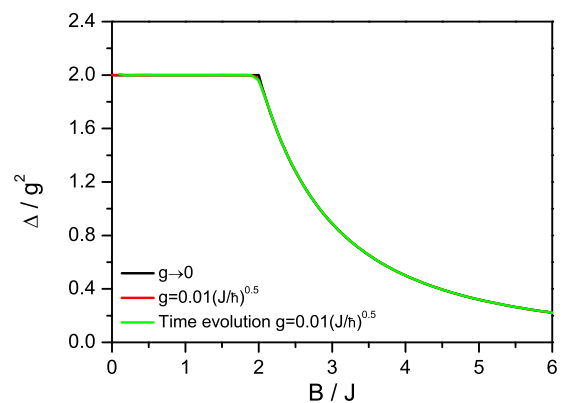}
\caption{(Color online) The ADRs $\Delta$ [see Eq.~\eqref{final result}]
 of the XY chain \eqref{XY chain} for $\gamma=1$ for
  $g=0.01\kla{J/\hbar}^{0.5}$ and $g\rightarrow 0$ (result of perturbation
  theory) as a function of the magnetic field $B$ are compared with the
  late-time decoherence rates extracted from Fig.~\ref{dynamics}. The
  agreement between the ADR and the late-time
  decoherence rate shows the validity of our calculations for finite times.}
\label{comparison}
\end{figure}

The final result for the ADR \eqref{final result} is valid in the limits $N
\rightarrow \infty$ and $g \rightarrow 0$. In this section we perform a
numerical diagonalization of the Lindblad master equation superoperator
$\mathcal{S}$ to compare the analytic result with the values for finite $N$
and $g$. Furthermore, we extract the ADR from a simulation of the system
dynamics and compare it with our prediction.

In Fig.~\ref{finite coupling} we present the ADR for finite coupling strengths
$g$. For $g^2\le 0.01J/\hbar$ the result of perturbation theory is in
excellent agreement with the numerical diagonalization of the Lindblad master
equation superoperator. Deviations are strongest at small magnetic fields
for which the finite $g$ is no longer a small
  perturbation. The non-analytic behavior at the critical field value $B/J=2$
  is clearly visible. The additional structure in the ADR for finite $g$ and
  small $B/J$ arises from level crossings in the spectrum of the
  Liouvillian. At $B=0$ the steady state becomes highly degenerate. The ADR
  (the largest \emph{non-zero} real part) jumps to a finite value indicating a
  finite gap above the steady-state manifold.

We show the ADR
$\Delta$ for different (finite) system sizes in Fig. \ref{finite size}. Even in
small systems with $N=10$ spins the same qualitative behavior is found as in
thermodynamic limit, i.e., the ADR signals the quantum
phase transition in the system at $B=2J$. However, finite values of $g$ and
$N$ lead to a smearing out of the phase transition.

We have defined the ADR through a diagonalization of the master equation,
trying to describe the long-time dynamics of the system. To demonstrate the
deep relation between $\Delta$ and the dissipative dynamics, we extract the
decoherence rate from a dynamical calculation (see Fig. \ref{dynamics}). Here
we start from the ground state of the system and study the decay of the
magnetization in time after the system is brought into contact with a
Markovian bath. In this example the exponential decay expected after long
evolution times is nicely visible. In Fig. \ref{comparison} we compare the
extracted decay rates for different magnetic fields with the result of the
diagonalization. We find an exact agreement with the ADR numerically
calculated with the same finite parameters.

We can calculate the ADR for the XY chain for general
values of $\mu$ and $\nu$ in a similar way. In the spin picture the Lindblad
operators are
\begin{gather}
 \mathbf{L}^\alpha_z=g\mu\sigma_\alpha^z=g\mu\frac{i}{2}\left[c_{\alpha,1},c_{\alpha,0}\right],\\
 \mathbf{L}^\alpha_x=g\nu\sigma_\alpha^x\sigma_{\alpha+1}^x=g\nu\frac{i}{2}\left[c_{\alpha,0},c_{\alpha+1,1}\right].
\end{gather}
We find for the constants in Eq. \eqref{full decoherence rates} in the case
$\gamma=1$
\begin{gather}
 \epsilon_z/\mu^2=\begin{cases}
                \frac{1}{2} & B\le 2J\\
		1-\frac{1}{2}\kla{\frac{2J}{B}}^2 & B\ge 2J,
               \end{cases}\\
\epsilon_x/\mu^2=\begin{cases}
                1-\frac{1}{2}\kla{\frac{B}{2J}}^2 & B\le 2J\\
		\frac{1}{2} & B\ge 2J,
               \end{cases}\\
\epsilon/(\mu\nu)=\begin{cases}
                -\frac{B}{2J} & B\le 2J\\
		-\frac{2J}{B} & B\ge 2J.
               \end{cases}\\
\end{gather}
In the symmetric case $\mu=\nu$, the ADR is constant
$\Delta=-4g^2\mu^2$. However, the next larger decoherence rate changes
non-analytically:
\begin{equation}
 \Lambda_{-}=\begin{cases}
              -2g^2\mu^2\kla{3+\kla{\frac{B}{2J}}^2}&\text{if } B<2J\\
	      -2g^2\mu^2\kla{3+\kla{\frac{2J}{B}}^2}&\text{if } B>2J.
             \end{cases}
\end{equation}

\section{Experimental Realization}
\label{Experiment}
We now discuss an experiment suited for the measurement of the ADR in spin
systems. The quantum simulation of spin systems with trapped ions was proposed
in \cite{Porras04}, where the spin degree of freedom is represented by two
hyperfine levels. The magnetic field can be simulated either by directly
driving Rabi oscillations of the hyperfine transition or with
position-independent Raman transitions induced by suitably aligned lasers. The
spin-spin interaction is mediated via motional degrees of
freedoms. State-dependent optical dipole forces (compare with state-dependent
optical lattices) are generated by coupling the two hyperfine levels to
electronically excited states with off-resonant laser beams. These dipole
forces change the distance and consequently the Coulomb repulsion between two
ions dependent on their internal states. This state-dependent Coulomb
repulsion can be designed to give the required spin-spin interaction. The spin
state can be measured by fluorescence imaging of the ions.

In this way the quantum Ising chain \cite{Friedenauer08,Schmitz09} and
frustrated Ising models \cite{Kim10} have been realized in recent
experiments.  In these experiments the ions were first cooled to their
zero-point motional ground state and optically pumped into a certain
spin configuration representing the ground state of the system without
spin-spin interactions. Then the spin-spin interactions were
adiabatically increased such that the system underwent a phase
transition. Finally, it was checked that the final state represented
the ground state of the simulated Hamiltonian. A large non-critical 2d
Ising system has been simulated with ions in a Penning trap
\cite{Britton2012}.  In the digital approach to quantum simulation
with trapped ions, the elements of a general toolbox including
Hamiltonian and dissipative dynamics have been demonstrated
\cite{Lanyon2011,Barreiro2011}.

We describe in the following how to extend analog quantum simulation
to include an incoherent evolution. The Lindblad master equation
\eqref{Lind1} with Hermitian Lindblad operators
$\mathbf{L}^\alpha=g\sigma_z^\alpha$ (see Sec.~\ref{quadratic with
  Hamiltonian}) can be realized by introducing fluctuations of the
simulated magnetic field $B^\alpha(t)=B^\alpha+\delta B^\alpha(t)$
\cite{Marquardt08} as shown in the following. The local magnetic
fields $\delta B^\alpha(t)$ should be uncorrelated between different
sites $\overline{\delta B^\alpha(t_1)\delta
  B^\beta(t_2)}=\delta_{\alpha\beta}\overline{\delta
  B^\alpha(t_1)\delta B^\alpha(t_2)}$. We restrict our derivation to a
single Lindblad operator without loss of generality. Let, for example,
$\delta B(t)$ constitute a Gaussian stochastic process of zero mean
$\overline{ \delta B(t)}=0$ with the time-correlations
\begin{equation}
\overline{\delta B(t_1) \delta B(t_2)}=\frac{ \overline{\delta B^2} }{
  \sqrt{2\pi} } \exp[-\frac{(t_1-t_2)^2}{2T^2}].
\end{equation}
The correlation time $T$ has to be much shorter than every process in the
system (Markovian limit), i.e., $\|\bcH\|T<\omega T\ll 1$, with the
spectral width $\omega$ of the Hamiltonian (difference between largest and
smallest eigenvalue) and the superoperator $\bcH$ from
Eq. \eqref{superoperator}. The averaged density matrix evolves like
$\ket{\rho(t)}=\overline{\bcU\kla{t}}\ket{\rho(0)}$, where the bar
denotes the statistical average over the fluctuating magnetic field. The time
evolution operator $\overline{\bcU\kla{t}}$ consists of contributions
from $\bcH$ and
\begin{equation}
 \bcV(t)=\frac{\delta B(t)}{\hbar}\bcV=-\frac{i\delta B(t)}{\hbar}\kla{\mathbf{V}\otimes\mathbf{1}-1\otimes\mathbf{V}^\text{T}}.
\end{equation}
with $\mathbf{V}=\sigma_z$. We can evaluate the statistical average of the time evolution operator in the interaction picture for the superoperators 
\begin{widetext}
\begin{align}
 \overline{\bcU\kla{t}}&=e^{\bcH t}\overline{ \mathcal{T}\exp\kla{\int_0^t d\tau e^{-\bcH\tau}\bcV(\tau)e^{\bcH\tau}}}\nonumber\\
&= e^{\bcH t}\sum_{n=0}^\infty \overline{ \int\limits_{t\ge t_1\ge\dots\ge t_n\ge 0} dt_1\dots dt_n e^{-\bcH t_1}\bcV(t_1)e^{\bcH t_1}\cdots e^{-\bcH t_n}\bcV(t_n)e^{\bcH t_n}}\nonumber\\
& =e^{\bcH t}\sum_{m=0}^\infty \kla{ \frac{1}{\hbar^2}\int_0^\infty
    \overline{ \delta B(0) \delta B(\tau)} d\tau }^{m} \cdot
\int\limits_{t\ge t_1\ge\dots\ge t_{m}\ge 0} dt_1\dots dt_{m}  e^{-\bcH t_1}\bcV^2e^{\bcH t_1}\cdots e^{-\bcH t_{m}}\bcV^2e^{\bcH t_{m}}\nonumber\\ 
&=e^{\bcH t}\sum_{m=0}^\infty \kla{\frac{\overline{ \delta B^2}}{\hbar^2}\cdot\frac{T}{2}}^{m} \cdot
\int\limits_{t\ge t_1\ge\dots\ge t_{m}\ge 0} dt_1\dots dt_{m}
e^{-\bcH t_1}\bcV^2e^{\bcH t_1}\cdots e^{-\bcH t_{m}}\bcV^2e^{\bcH t_{m}}\nonumber\\
\overline{\bcU\kla{t}}&=\exp\kla{\bcH t +\frac{1}{2}\frac{\overline{ \delta B^2} T}{\hbar^2} \bcV^2t}
\end{align}
\end{widetext}
with the time ordering operator $\mathcal{T}$. Between the second and the
third line, we keep only even summation indices $m=2n$ (zero mean Gaussian
process), evaluate the statistical average at adjacent times
$t_{2n-1}-t_{2n}\le T$ (correlation time $T$; that only
  adjacent times need to be considered is a consequence of time-ordering, the
  Gaussian factorization of higher-order correlations, and
  the very short correlation times), and neglect the terms
$\exp[\bcH(t_{2n-1}-t_{2n})]\ll 1$ (Markovian limit). In summary,
we have shown that the described fluctuations of the magnetic field generate
Markovian dynamics [see Eq.~\eqref{Lindblad3}] with Lindblad operators
$\mathbf{L}^\alpha=g\sigma_z^\alpha=g\mathbf{V}$ and decoherence strength
\begin{equation}
 g^2=\frac{\overline{\delta B^2}T}{\hbar^2}.
\end{equation}
In the case of the anisotropic XY chain [see Eq.~\eqref{XY chain}], the
correlation time $T$ is bounded by the width of the single particle excitation
spectrum $T^{-1}\gg \text{max}\kla{4B/\hbar,8J/\hbar}$. In the recent
experiment \cite{Friedenauer08} $2J/\hbar\approx B/\hbar=2\pi \times
\unit{4.4}{\kilo\hertz}$ was used, but experimentally available laser
intensities allow $2J/\hbar\approx B/\hbar\approx 2\pi \times
\unit{40}{\kilo\hertz}$. We propose to create fluctuations of the magnetic
field with frequency $T^{-1}=2\pi\times \unit{1.6}{\mega\hertz}$ and variance
$\overline{{\delta B^2}}/\hbar^2=(0.2B/\hbar)^2\approx (2\pi\times
\unit{8}{\kilo\hertz})^2$. This would result in the decoherence strength
$g^2\approx \unit{2\cdot10^{-3}}{\joule}/\hbar$ and would require coherence
times of order $2\pi/g^2\approx \unit{25}{\milli\second}$. These coherence
times can in principle be achieved in systems of trapped ions
\cite{Wineland98}.

\section{Conclusion}
\label{Conclusion}
We have investigated the dynamics of open quantum systems with regard to their
steady states and asymptotic decay. We have shown that insight into different
phases can be gained by spectral analysis of the Liouvillian in analogy to how
the spectrum of the Hamiltonian reveals critical behavior in zero-temperature
quantum phase transitions.

To illustrate this point we have analyzed in detail the Liouvillian of
open fermionic systems under a translationally invariant, quadratic
Hamiltonian, coupled to a Markovian bath. We treat master equations
with linear or quadratic and Hermitian Lindblad operators. In both
cases, the master equation leads to a closed equation for the
CM from which the steady-state CM and
the rates at which it is approached can be obtained exactly (see also
\cite{EiPr11} for an elegant and comprehensive treatment of both
fermionic and bosonic linear open systems and their critical
properties and \cite{Znidaric11} for a detailed study of transport in
spin chains under dissipation and dephasing). These results apply as
well to a large class of 1d spin systems that can be mapped to
quasifree fermions by a Jordan-Wigner transformation.  We have
proposed an experimental realization of this quantum simulation with
trapped ions. Numerical calculations show that our results for the
weak decoherence limit do apply to such finite systems.

We have focused on the limit of weak decoherence ($g\to0$) and shown how to deduce
information about critical points from the spectrum of the Liouvillian. In
particular, the ADR $\Delta$, i.e., the smallest non-zero eigenvalue of the
Liouvillian, can serve an an indicator of phase transitions even if the steady
state of the system is trivial and steady-state expectation values thus cannot
yield such information (as in the case of Hermitian Lindblad
operators). Depending on the decoherence process considered, the critical
point can be reflected in the spectrum of the system's Liouvillian in the form
of a closing gap ($\Delta\to0$), a degeneracy of $\Delta$ or non-analytic
behavior of $\Delta$. These results are summarized in Table~\ref{tab:1}.

    \begin{table}[h]
  \centering
\caption{(Color online) Different dissipative systems studied, characterized by their
  Lindblad operators and Hamiltonian $H$. Relevant properties of the
  ADR $\Delta$ and the steady state are
  listed. $x_c$ denotes critical points of the Hamiltonian $H$.}
  \label{tab:1}
\begin{tabular}{|p{2.8cm}|p{3.3cm}|p{1.9cm}|}
\hline
\multicolumn{1}{|c|}{Lindblad Op}& \multicolumn{1}{|c|}{ADR and Gap} &
\multicolumn{1}{|c|}{Steady State} \\
\hline
\multicolumn{3}{|c|}{Hamiltonian $H=0$}\\
\hline
$\mu a_\alpha, \nu a_\alpha^\dag$ &   gapped, no p.t. & thermal\\ 
\hline
$\mu a_\alpha+\nu a_{\alpha+1}^\dag$ &   gap closes @ $\mu=\nu$, no p.t.&  paired
\\  
\hline
$\frac{i\mu}{2}[c_{\alpha,0},c_{\alpha,1}]$,\newline$\frac{i\nu}{2}[c_{\alpha+1,0},c_{\alpha,1}]$&   degenerate @ $\mu=\nu$     & $\propto\id$             \\
\hline
\multicolumn{3}{|c|}{Hamiltonian $H\not=0$: transl. invariant,
  critical at $x_c$}\\
\hline
$\mu a_\alpha, \nu a_\alpha^\dag$& degenerate @ $x_c$  &  $\langle a^\dag
a\rangle$ non-analytic~@$x_c$             \\ 
\hline
$\frac{i\mu}{2}[c_{\alpha,0},c_{\alpha,1}]$,\newline$\frac{i\nu}{2}[c_{\alpha+1,0},c_{\alpha,1}]$ &   non-analytic @ $x_c$    & $\propto\id$             \\ 
\hline
\end{tabular}
\end{table}

With this work we suggest the possibility of detecting certain system
properties through an observation of the decoherent dynamics: phase
transitions in closed systems can be reflected in non-analytic changes of the
ADR \cite{Znidaric11,Horstmann2011,HMF12}. More generally, since the ADR and
other decay rates represent physical properties of the system, such
non-analyticities can be seen as signature of a transition to a different
dynamical regime. This suggests to study the phase diagram of steady-state
correlation functions $\langle A(t)B(t')\rangle$ which will reflect these
dynamical transitions.

\begin{acknowledgments}
  The authors thank M.~M. Wolf and T. Roscilde, BH thanks M. Lubasch, L. Mazza,
  M.~C. Ba\~nuls, N. Schuch, and A. Pflanzer for fruitful discussions. The
  authors would like to acknowledge financial support by the DFG within the
  Excellence Cluster Nanosystems Initiative Munich (NIM), and the EU project
  MALICIA under FET-Open grant number 265522.
\end{acknowledgments}

\bibliographystyle{apsrev4-1}
\bibliography{Dissi}

\end{document}